\documentclass[aps,pra,reprint,groupedaddress]{revtex4-1}

\usepackage{xcolor} 
\usepackage{etoolbox,xpatch} 
\usepackage{graphicx} 
\usepackage{bm,upgreek,mathrsfs,amssymb,mathtools,physics} 

\usepackage{hyperref} 
\usepackage[capitalize]{cleveref}  
\definecolor{darkBlue}{rgb}{0.0235,0.27,0.678} 
\hypersetup{colorlinks=true,linkcolor=darkBlue,urlcolor=darkBlue,citecolor=darkBlue} 

\begin{document}


\title{General framework for treating generation, propagation, and polarization of luminescence in anisotropic media}


\author{Shane Nichols}
\author{Melissa Tan}
\author{Alexander Martin}
\author{Emily Timothy}
\author{Bart Kahr}\email[]{bart.kahr@nyu.edu}
\affiliation{New York University, Department of Chemistry, New York, NY 10003}


\date{\today}

\begin{abstract}
Complete polarimeters deliver the full polarization transfer matrix of a medium that relates input polarization states to output polarization states. In order to interpret the Mueller matrix of a luminescent medium at the emission frequency, accountings are required for polarization transformations of the medium at the excitation frequency, the light scattering event, and the polarization transformations at the emission frequency. A general framework for this kind of analysis is presented herein. The fluorescence Mueller matrix is expressed as a product of Mueller matrices of the medium at the excitation and emission frequencies and a scattering matrix, integrated over path length. The Stokes vector for the incident light, evolving according to the Mueller matrix of the medium at the excitation frequency, is multiplied by the scattering matrix to give a Stokes vector of the emitted light along the propagation direction. The scattering matrix is derived from an incoherent orientation ensemble average over a phenomenological scattering tensor that embodies intrinsic molecular scattering proprieties, and dynamical processes that occur during the excited state. The general framework is evaluated for three anisotropic materials carrying luminescent dye molecules including the following: a chiral fluid, a stretched polymer film, and a chiral, biaxial crystal. In the latter case, the most complex, the Mueller matrix was collected in conoscopic illumination and the fluorescence Mueller matrix, mapped in k-space, was fully simulated by the strategy outlined above. Luminescence spectroscopy has typically stood apart from the developments in polarimetry of the past two generations. This need not be indefinite.
\end{abstract}

\pacs{}

\maketitle

\section{Introduction}
Polarization properties of fluorescence and the mathematical foundations that underly modern polarimetry have a common origin. In his comprehensive memoir "On the change of refrangibility of light" \cite{Stokes:1852aa}, Stokes coined the term `fluorescence' \textit{and} introduced the so-called Stokes parameters that describe the polarization state of light. He observed that fluorescence from solutions is unpolarized while that from crystals is often polarized. Soleillet later showed that solutions will emit polarized light if excited with polarized light and if the analyte rotates slowly on the time scale of the excited state \cite{Soleillet:1929aa}. He presented four equations that connect the Stokes parameters of the incident excitation to the scattered emission, which depend on the scattering angle and a quantity called fluorescence anisotropy. Soleillet was also the first to invoke what is now called the differential Mueller matrix formalism \cite{Arteaga:2017aa} and used it to consider the effects of anisotropy on the exciting beam \cite{Arteaga:2012ac}. Perrin further developed fluorescence polarization to study anisotropic rotational diffusion \cite{Perrin:1928aa}, and considered scattering by optically active and oriented systems in the context of the Stokes parameters \cite{Perrin:1942aa}. Despite this history, the field of fluorescence polarization has developed spectacularly \textit{outside} of the Mueller-Stokes framework. 

Fluorescence polarization has since become an important tool in biophysics \cite{Chen:1999aa,Burke:2003aa,Owicki:2000aa,Forkey:2003aa,Valeur:2012aa}. It has been used to detect binding of small molecules to large biomolecules, and to characterize liquid crystals \cite{Johansson:1980aa,Dozov:1980aa,Naqvi:1980aa,Subramanian:1982aa,Zannoni:1983aa,Bauman:1988aa,Wolarz:1992aa,Hashimoto:1993aa,Schartel:1999aa}, Langmuir-Blodgett films \cite{LeGrange:1988aa,Edmiston:1996aa}, sheared melts \cite{Bur:1991aa}, and biological membranes \cite{Fuchs:1975aa,Lentz:1976aa,Jahnig:1979aa}. Additionally, polarized fluorescence microscopy has shown the local orientation and dynamics in cellular membranes and vesicles \cite{Axelrod:1979aa,Vrabioiu:2006aa,Gasecka:2009aa,Jameson:2010aa,Kampmann:2011aa,Owen:2012aa,Kress:2013aa,Zhanghao:2016aa}. Yet, the great shortcoming of the development of fluorescence polarization outside of polarimetry is the failure to explicitly consider polarization transformations in the excitation beam and in the emission beam, save elementary considerations of linear dichroism at the excitation wavelength. While membranes and thin-films have path lengths that likely do not warrant such considerations, liquid crystals, melts, and stretched films most likely do. Accountings of polarization in the Mueller matrix formalism have been reported \cite{Shindo:1992aa,Harada:2012aa}, but with severe approximations.

Recently, researchers in biomedicine have taken a keen interest in implementing Mueller matrix fluorimetry for the characterization of tissues \cite{Soni:2013ab,Alali:2015aa,Ushenko:2015aa,Saha:2015aa}. In such cases, developing models of light generation and propagation are less important than establishing robust fingerprints to distinguish one tissue from another, usually for disease recognition. Fluorescence Mueller matrix (FMM) measurements have also been used to characterize scattering from collagen fibers \cite{Mazumder:2017aa}, polycrystalline films of biological origin \cite{Ushenko:2016ab}, polymeric nanoparticles \cite{Satapathi:2014aa}, and supramolecular assemblies \cite{Maji:2017aa}. Many of these works and others \cite{Arteaga:2012ac,He:2013aa} suggest various phenomenological parameters to quantify effects of propagation and scattering, but are not firmly rooted in the differential Mueller matrix formalism. Our aim here is to develop robust methods to handle the effects of propagation at both the excitation and emission wavelengths, and to substantiate the methods proposed with simple, illustrative measurements which correspond to practical optical systems. The three examples include a fluorophore in an optically active isotropic medium (glucose), a fluorophore in an anisotropic medium (a stretched polymer film), and a fluorophore in a bianisotropic medium, (a dyed, chiral crystal).

\section{Theory}

While the differential Mueller matrix formulation was introduced nearly 40 years ago \cite{Azzam:1978ab} and has since been well studied \cite{Ossikovski:2014aa,Devlaminck:2014aa,Ossikovski:2011aa}, its form in the presence of luminescent sources has not been considered. In a continuous medium with luminescent sources throughout, the evolution of the Stokes vector at the emission wavelength $\bm s_\mathrm{e}(z)$ at a point $z$ along the propagation direction is governed by the nonhomogeneous differential equation
\begin{equation}\label{eq:dS}
\mathrm{d}\mathbf s_\mathrm{e}/\mathrm{d}z = \mathbf m_\mathrm{e} \bm s_\mathrm{e} + \bm s_s
\end{equation}
where $\bm s_s$ is a $z$-dependent source term that represents the Stokes vector of light emitted at $z$ that is directed along the propagation direction, and $\mathbf m_\mathrm{e}$ is the differential Mueller matrix at the emission wavelength. Using the method of variation of parameters, the general solution to \cref{eq:dS} is
\begin{equation}\label{eq:SzFull}
	\bm s(z) = \mathbf M_\mathrm{e}(z) \mathbf M^{-1}_\mathrm{e}(0)\bm s(0) +
	\mathbf M_\mathrm{e}(z)\int_0^z  \mathbf M^{-1}_\mathrm{e}(z')\bm s_s(z') \ \mathrm{d}z'
\end{equation}
where $\mathbf M_\mathrm{e}$ is the Mueller matrix describing propagation at the emission wavelength and is the solution to the corresponding homogeneous system, i.e., it satisfies $\mathrm{d}\mathbf M_\mathrm{e}/\mathrm{d}z = \mathbf m_\mathrm{e}\mathbf M_\mathrm{e}$. Physically, the first term on the right of \cref{eq:SzFull} propagates incident light at the emission wavelength through the medium whereas the second term accounts for the generation and propagation of luminescence within the medium. The integral in \cref{eq:SzFull} represents an incoherent superposition and is only appropriate for light generation processes in which the phases of light emitted from different points in the medium are random, e.g., spontaneous emission and vibrational Raman scattering \cite{Barron:2004aa}. Coherently driven processes, such as nonlinear mixing and stimulated emission, should be described by an analogous formulation of \cref{eq:dS} in terms of the differential Jones formalism \cite{Jones:1948aa}. 
 
Evaluation of \cref{eq:SzFull} requires a model for $\bm s_s$. In the case of fluorescence (or any kind of spontaneous photoluminescence), the source term is derived from the excitation field, described by its incident Stokes vector $\bm s_\mathrm{x}(0)$ and the Mueller matrix $\mathbf M_{\mathrm{x}}(z)$ by which it propagates, according to 
\begin{equation}\label{eq:excitationToSource}
\bm s_s(z) = \mathbf S(z) \mathbf M_{\mathrm{x}}(z)\bm s_\mathrm{x}(0).
\end{equation}
$\mathbf S$ is a scattering matrix that relates the instantaneous value of $\bm s_\mathrm{x}(z)$ to the net polarization of fluorescence that it generates. To obtain the overall FMM $\mathbf F$ that connects the incident Stokes vector at the excitation wavelength to the outgoing Stokes vector at the emission wavelength, we substitute \cref{eq:excitationToSource} into \cref{eq:dS}, introduce the path-length $d$ over which fluorescence is integrated, and drop $\bm S_\mathrm{x}(0)$, giving
\begin{equation}\label{eq:Sz}
\mathbf F(d) = \mathbf M_{\mathrm{e}}(d)\int_0^d \mathbf M^{-1}_{\mathrm{e}}(z) \mathbf S(z) \mathbf M_{\mathrm{x}}(z)  \ \mathrm{d}z.
\end{equation}
Note that the first term on the right of \cref{eq:SzFull} has vanished because we assume that no light enters the medium at the emission wavelength. The form of \cref{eq:Sz} is valid for media in which $\mathbf m_\mathrm{e}$ may be continuously varying in $z$, for example a twisted nematic liquid crystal, but hereafter we will consider only homogeneous media, where $\mathbf m_\mathrm{e}$ and $\mathbf S$ are $z$-independent. Such media have the simplifying properties $\mathbf M(a) = \exp(a\mathbf m)$, and $\mathbf M(a)\mathbf M(b)^{-1} = \exp((a-b)\mathbf m)$ that permit rewriting \cref{eq:Sz} as
\begin{equation}\label{eq:SzHomo}
\mathbf F(d) = \int_0^d \exp((d-z)\mathbf m_\mathrm{e}) \mathbf S \exp(z\mathbf m_\mathrm{x})  \ \mathrm{d}z,
\end{equation}
in which $ \mathbf m_\mathrm{x}$ is the differential Mueller matrix at the excitation wavelength. \Cref{eq:SzHomo} is intuitive in that the integrand expressly gives the FMM for light generated at $z$; the excitation propagates to $z$ according to $\exp(z\mathbf m_\mathrm{x})$, generates fluorescence according to $\mathbf S$, which then propagates over the remaining path length according to $\exp((d-z)\mathbf m_\mathrm{e})$. The overall FMM is then the integral over the path length. 

\begin{figure}
\centering
\includegraphics[width=0.9\linewidth]{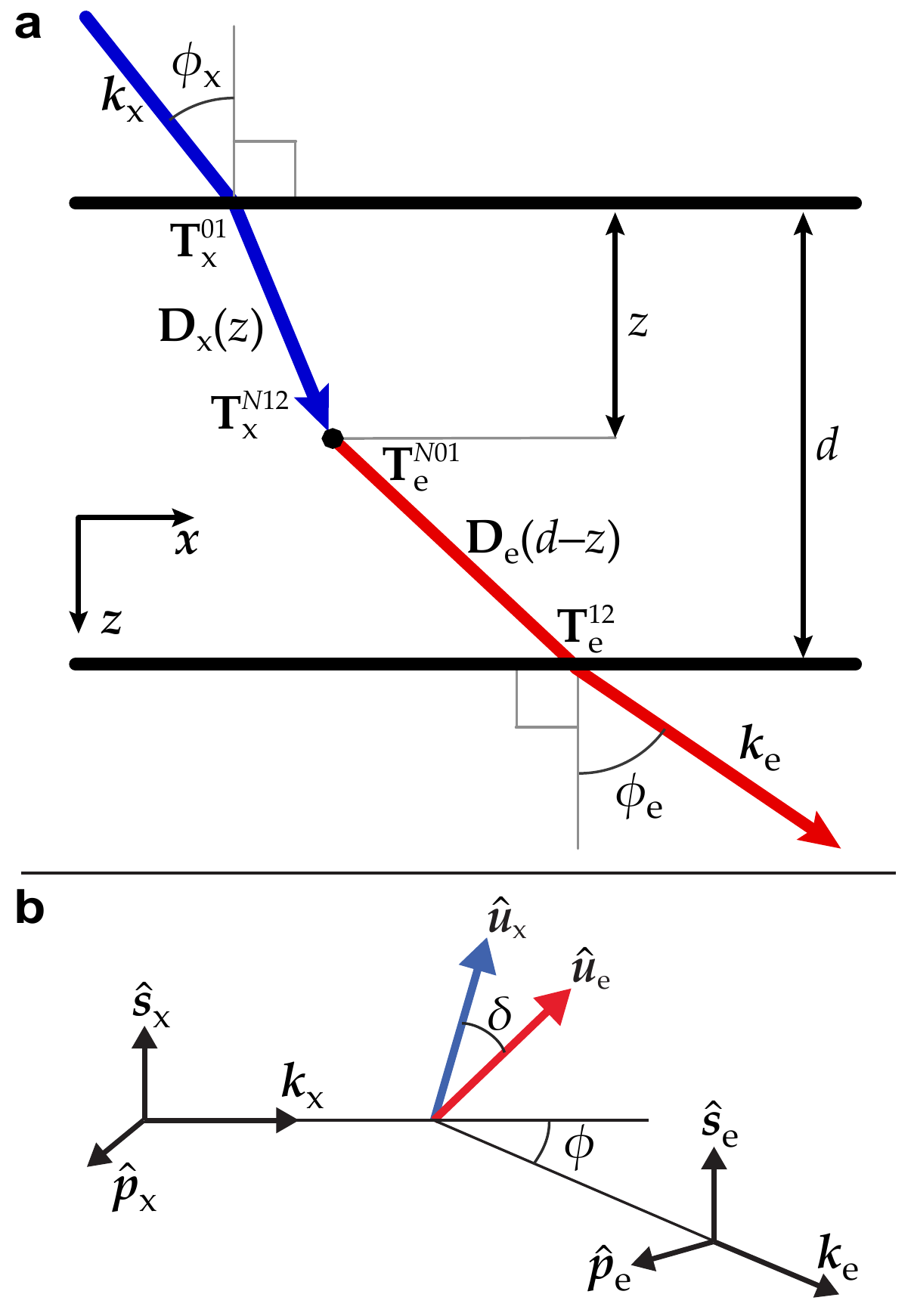}
\caption{Scattering coordinate systems. (a) Diagram of the propagation Jones matrices used to incorporate boundary conditions. (b) A general scattering coordinate system in which the excitation and emission wave vectors are, respectively, $\bm k_\mathrm{x}$ and $\bm k_\mathrm{e}$. The electric fields are projected onto directions in the plane ($\hat{\bm p}_i$) and out of the plane ($\hat{\bm s}_i$) of scattering. Dipole scattering can be described by an absorption dipole $\hat{\bm u}_\mathrm{x}$ and an emission dipole $\hat{\bm u}_\mathrm{e}$.  }
\label{fig:scattering}
\end{figure}

Incorporation of boundaries requires a more general formulation than \cref{eq:SzHomo}. Rigorous accountings of light generation in arbitrary homogeneous media typically begin with incorporating source terms into Maxwell's equations. Solutions are given in terms of dyadic Green's functions, in which the net electric field in the exit medium is written as a volume integral \cite{Bladel:1991aa,Pazynin:2016aa,Huang:2017aa}. We question, however, whether this approach is appropriate for luminescence because an integral over field amplitudes is fundamentally a coherent superposition, analogous to casting \cref{eq:SzFull} as Jones matrices. Moreover, deriving the requisite Green's functions is usually a formidable task \cite{Mackay:2010aa}. Here we introduce a simple, approximate method to incorporate interfaces that is based on the reduction of propagation in plane parallel media to partial waves \cite{Postava:2002aa,Nichols:2015aa}.

Consider a medium bounded by plane parallel interfaces normal to $\bm z$ at positions $z=0$ and $z=d$, as shown in \cref{fig:scattering}. The excitation wave vector makes an angle $\phi_\mathrm{x}$ with the substrate normal and is contained in the $xz$ plane. In general, the incident field excites in the slab two forward and two backward propagating plane waves. If we ignore reflections from inside the slab, the excitation propagation is fully described by three $2\times2$ Jones matrices; $\mathbf T_\mathrm{x}^{01}$ maps the electric field in the incident ambient medium, described in the $sp$ basis, to the amplitudes of the forward propagation eigenwaves within the layer; $\mathbf T_\mathrm{x}^{12}$ maps the forward eigenwaves inside the layer to the $sp$ basis in the ambient exit medium; $\mathbf D_\mathrm{x}$ is the diagonal matrix that propagates the eigenwaves across the layer and has elements $D_{jj} = \exp(-ik_0dn_j)$ where $n_1$ and $n_2$ are the complex refractive indices at which the eigenwaves propagate and $k_0=2\pi/\lambda$ is the free-space wavenumber.

Propagation Jones matrices $\mathbf T_\mathrm{e}^{01}$, $\mathbf T_\mathrm{e}^{12}$, and $\mathbf D_\mathrm{e}$ are computed for the emission wavelength in a relative coordinate system where the emission wave vector in the exit ambient medium is contained in the $x'z$ plane and makes an angle $\phi_\mathrm{e}$ with the substrate normal. Optical tensors of the medium are transformed accordingly. In this way, equations given elsewhere \cite{Nichols:2015aa} to obtain the Jones matrices can be used without modification. The coordinate systems for the excitation and emission fields are related by a rotation of $\theta$ about the $z$ axis. 

To describe fluorescence scattering inside the medium, the excitation and emission fields must be brought from their respective eigensystems into a common coordinate system. One option is to work within the general coordinate system of the Berreman transfer matrices \cite{Berreman:1972aa}, from which the propagation Jones matrices are derived, but we achieved nearly identical results by simply normalizing $\mathbf T^{01}$ and $\mathbf T^{12}$ such that the reflection losses that they innately describe are mitigated, and their effect is only to map the fields into an $sp$ basis. The normalized matrices are defined by $\bm T_i^{N12} = \bm T_i^{12}/\norm*{\bm T_i^{12}}$ and $\bm T_j^{N01} = \bm T_j^{01}/\norm*{\bm T_j^{01}}$ where $i$ is a row index and $j$ is a column index. We can now define the overall Jones matrices that handle propagation before and after the scattering event as
\begin{subequations}\label{eq:JwithBoundaries}\begin{gather}
\mathbf J_\mathrm{x}(z) =\mathbf T_\mathrm{x}^{N12}\mathbf D_\mathrm{x}(z)\mathbf T_\mathrm{x}^{01}, \\
\mathbf J_\mathrm{e}(d-z) = \mathbf R(\theta)\mathbf T_\mathrm{e}^{12}\mathbf D(d-z)\mathbf T_\mathrm{e}^{N01}\mathbf R(-\theta)
\intertext{where}
\mathbf R(\theta) = \mqty[\cos\theta & \sin\theta \\ -\sin\theta & \cos\theta]
\end{gather}\end{subequations}
is a rotation matrix that brings the emission field into the global coordinate system. Both Jones matrices are converted to Mueller matrices by $\mathbf M_n = \mathbf A(\mathbf J_n\otimes \mathbf J_n^*)\mathbf A^{-1}$, where $\otimes$ is the Kronecker product, and
\begin{equation}\label{eq:Amatirx}
\mathbf A = \begin{bmatrix}
		1&0&0&1\\1&0&0&-1\\0&1&1&0\\0&i&-i&0
	\end{bmatrix}.
\end{equation}
Finally, the Mueller matrices are inserted into \cref{eq:Sz} to obtain the FMM.

We now turn our attention to the scattering matrix $\mathbf S$. To provide the most general results, we make no assumptions as to the scattering geometry in its derivation. Properties of the fluorophores such as electronics, rotational dynamics, and energy transfer are embodied in the scattering events that convert excitation to emission. However, $\mathbf S$ is not a standalone property of the medium but corresponds to a particular scattering geometry in which $\bm s_\mathrm{x}$ and $\bm s_\mathrm{e}$ propagate along arbitrary directions described by the respective wave vectors $\bm k_\mathrm{x}$ and $\bm k_\mathrm{e}$. As shown in \cref{fig:scattering}, the electric fields of both rays are projected onto unit vectors in ($\hat{\bm p}$) and out ($\hat{\bm s}$) of the scattering plane. In terms of this geometry, we introduce a $9\times9$ matrix $\mathbf Q$ that contains the three-dimensional scattering properties of the medium, from which $\mathbf S$ is extracted according to
\begin{multline}\label{eq:QtoS}
\mathbf S = \mathbf A\bigl(\mqty[ \hat{\bm p}_\mathrm{e}{\otimes}\hat{\bm p}_\mathrm{e} &
				   \hat{\bm p}_\mathrm{e}{\otimes}\hat{\bm s}_\mathrm{e} &
				   \hat{\bm s}_\mathrm{e}{\otimes}\hat{\bm p}_\mathrm{e} &
				   \hat{\bm s}_\mathrm{e}{\otimes}\hat{\bm s}_\mathrm{e}]^{\mathrm T}
		    \times \\
		    \mathbf Q 
		    \mqty[ \hat{\bm p}_\mathrm{x}{\otimes}\hat{\bm p}_\mathrm{x} &
		    	   \hat{\bm p}_\mathrm{x}{\otimes}\hat{\bm s}_\mathrm{x} &
		    	   \hat{\bm s}_\mathrm{x}{\otimes}\hat{\bm p}_\mathrm{x} &
		    	   \hat{\bm s}_\mathrm{x}{\otimes}\hat{\bm s}_\mathrm{x}]\bigr)\mathbf A^{-1}
\end{multline}
where each Kronecker product produces a 9-element column vector.

In essence, $\mathbf Q$ represents an ensemble average over all molecular scattering events that can occur in the medium. Herein we assume that the scattering process is electric-field mediated, thus a single scattering event can be described by a $3\times 3$ tensor $\mathbf P$ that relates the three-dimensional incident and scattered electric fields according to $\bm E_s = \mathbf P\bm E_i$ \cite{Barron:2004aa}. For elastic (coherent) scattering, an orientation ensemble of scatterers is described by direct integration over $\mathbf P$, however, because fluorescence is intrinsically incoherent, any ensemble average must be performed on second order moments \cite{Hingerl:2016aa,Ossikovski:2016aa,Nichols:2016ab}. In this case, $\mathbf Q=\expval{\mathbf P\otimes\mathbf P^*}$, which may be formally written as
\begin{equation}\label{eq:Qformal}
	\mathbf Q = \dfrac{\int_\Omega w(\Omega)\mathbf P\otimes\mathbf P^* \ \mathrm{d}\Omega}
		{\int_\Omega w(\Omega) \ \mathrm{d}\Omega}
\end{equation}
with $\mathrm{d}\Omega$ being a volume element of orientation space, and $w(\Omega)$ a weighting function giving the probability density of a fluorophore being in a particular orientation. There may be multiple tensors $\mathbf P$ that describe different possible scattering events, for example multiple electronic transitions at the excitation frequency, or emission from multiple excited states. Because fluorescence is incoherent, $\mathbf P$ describing each unique scattering event should be mapped to a tensor $\mathbf Q$, and the $\mathbf Q$'s summed. As molecules are much smaller than the wavelength of light, molecular scattering tends to be dipole-like \cite{Lakowicz:2008aa}, in which case $\mathbf P$ becomes the dyad $\mathbf P = \hat{\bm u}_\mathrm{e} \otimes \hat{\bm u}_\mathrm{x}$, where $\hat{\bm u}_\mathrm{x}$ and $\hat{\bm u}_\mathrm{e}$ are respectively unit vectors along the directions of the absorption and emission dipole moments of the fluorophore. For dipole scattering, circular polarization in the excitation field is inconsequential because the last row and column of $\mathbf S$ are zero for any orientation distribution of dipole scatterers. 

As a final note, we have neglected terms that relate to the absolute quantity of light emitted, such as the quantum yield, because polarimetric measurements are typically normalized by the overall light intensity. We have also neglected the usual $1/r^2$ dependence of radiation intensity under the assumption that the thickness of the specimen is much smaller than the distance between the specimen and the detector.

\section{Experimental}\label{app:experimental}
Measurements in \cref{sec:OAsol,sec:stretchedFilm} were performed with a previously described spectroscopic Mueller matrix polarimeter that uses four photoelastic modulators \cite{Arteaga:2012ad}. A broadband light source was used for transmission studies, while either a 441 nm laser (4 W, CW direct-diode, Lasertack GmbH) or a 532 nm laser (300 mW, CW DPSS, Lasermate) was used for fluorescence excitation. Low strain birefringence microscope objectives with 4$\times$ magnification and a numerical aperture of 0.1 were positioned on both sides of the specimen to focus the excitation and gather emission. Emission was separated from excitation using two high-contrast long-pass filters, and the emission was spectroscopically analyzed with a scanning grating monochromator. All measurements were performed at normal incidence in the forward scattering configuration using an instrument coordinate system in which $\hat{\bm p}_\mathrm{x} = \hat{\bm p}_\mathrm{e} = \bm x$ and $\hat{\bm s}_\mathrm{x}= \hat{\bm s}_\mathrm{e} = \bm y$.

The measurement in \cref{sec:SYEDS} was performed with a Mueller matrix microscope that uses continuous rotating retarders before and after the specimen to modulate the polarization state of light. Its operational principles are similar to others \cite{Azzam:1978aa,Chen:2004aa}, and its particular details are given elsewhere \cite{Nichols:2018aa}. A collimated 450 nm laser (3.5 W, CW direct-diode, Wicked Lasers) was used for excitation. Emission was gathered with a polarization grade 40$\times$ objective and passed through a 550-570 bandpass filter. The rear focal plane of the objective was imaged on a camera sensor to acquire $\bm k$-space (conoscopic) Mueller matrix images, similar to the device described in Ref. \cite{Arteaga:2017ab}.

\section{Results and discussion}
Three illustrative homogeneous media are shown to substantiate the equations of the previous section, and to examine some particular forms of $\mathbf S$. Knowledge of the materials are used to either predict $\mathbf S$, or reduce its form to a few parameters that may be fitted from the data. Application of \cref{eq:SzHomo} requires knowing $\mathbf m_\mathrm{x}$ and $\mathbf m_\mathrm{e}$. In a linear homogeneous medium, $\mathbf m$ has only seven parameters and takes the form
\begin{equation}\label{eq:diffMuellerJones}
	\mathbf m = \frac 1d \mqty[ 
		-A  & -L\!D & -L\!D' & C\!D \\
		-L\!D & -A & C\!B & L\!B' \\
		-L\!D' & -C\!B & -A & -L\!B \\
		 C\!D & -L\!B' & L\!B &  -A],
\end{equation} 
in which $L\!D$ and $L\!D'$ are, respectively, on and off axis linear dichroism, $L\!B$ and $L\!B'$ are, respectively on and off axis linear birefringence, $C\!D$ and $C\!B$ are respectively circular dichroism and circular birefringence, and $A$ is the mean absorbance \cite{Arteaga:2009ac}. These parameters are obtained either from an optical model of the medium in terms of the constitutive relations that solve Maxwell's equations, or from a measurement of the transmission Mueller matrix (TMM). In the latter case, however, care must be taken to ensure that the correct branch is selected from the multivalued function $\mathbf m = \ln(\mathbf M)/d$ \cite{Devlaminck:2014aa}. Although each branch maps to the same Mueller matrix of the homogeneous solution $\mathbf M = \exp(z \mathbf m)$, the same is not true for \cref{eq:SzHomo} due to the integral. Equations to obtain $\mathbf m$ from either optical models or transmission data are provided in \cref{app:1}.

\subsection{Isotropic distributions}  \label{sec:OAsol}

We previously considered the form of $\mathbf F$ for an isotropic distribution of fluorophores \cite{Arteaga:2012ac}, in which we recast Soleillet's classical results on fluorescence scattering \cite{Soleillet:1929aa} as the matrix
 \begin{equation}\label{eq:FIso}
 \mathbf S =  \mqty[
  a-b \sin ^2\phi & -b \sin ^2\phi  & 0 & 0 \\
  -b \sin ^2\phi  & b \left(1+\cos ^2\phi\right) & 0 & 0 \\
  0 & 0 & 2 b \cos \phi  & 0 \\
  0 & 0 &  &  c\cos\phi]
 \end{equation}
where $\phi$ is the scattering angle, and $a$, $b$, and $c$ are properties independent of $\phi$. We previously remarked that $c$ is related to optical activity \cite{Arteaga:2012ac}, but this statement was made in error. Because $c$ appears on the diagonal of $\mathbf S$, it does not induce asymmetry with respect to any state of polarization. Rather, parameters $b$ and $c$ respectively establish the extent to which linear and circular polarizations are transferred from the excitation to the emission. To relate these parameters to a tensor $\mathbf P$ that describes a scattering event, we evaluated \cref{eq:QtoS,eq:Qformal} with $w(\Omega)=1$ and obtained,
$a = (6 t_1+t_2+t_3)/10$,
$b = (3(t_2+t_3)/2 - t_1 )/10$, and
$c = (t_3-t_2)/4$,
where
$t_1 = \tr \bigl(\mathbf P\mathbf P^\dagger\bigr)$,
$t_2 = \tr \bigl(\mathbf P\mathbf P^\text{*}\bigr)$,
$t_3 = \tr(\mathbf P)\tr(\mathbf P^*)$, and $\mathrm{tr}$ denotes the trace. 
While this result suggests the possibility of $c\neq 0$, such an observation has yet to be made in any fluorescence experiment, to our knowledge, due to the dipole nature of molecular scattering, for which $c=0$. Higher multipoles in $\mathbf P$ do produce $c\neq 0$, however. For instance, if $\mathbf P$ takes the form of an arbitrary quadrupole moment, we find that $\mathbf S\propto \textrm{diag}(7, 1, 1, -2.5)$ in the forward scattering direction. Perhaps emissive nanoparticles may offer a route to obtain $c\neq0$. Such a system may be useful as it would exhibit emission that ``remembers'' the handedness of the excitation light.

For dipole scattering, where $\mathbf P = \hat{\bm u}_\mathrm{e}\otimes\hat{ \bm u}_\mathrm{x}$, the parameters $a$ and $b$ are more conveniently cast in terms of the angle $\delta$ between the absorption and emission dipole moments, i.e., $\cos\delta=\hat{\bm u}_\mathrm{e}\cdot\hat{\bm u}_\mathrm{x}$. It is known that $\delta\approx 0$ for typical fluorophores \cite{Lakowicz:2008aa}, but a molecule may behave as if $\delta$ is larger than its intrinsic value due to rotational diffusion of molecules while in the excited state. In that sense, the vectors $\bm u_\mathrm{e}$ and $\bm u_\mathrm{x}$ should be evaluated at the time of absorption and emission, respectively. Thus, rotational diffusion leads to a distribution in $\delta$, and the definitions
$a = \tfrac{1}{10}( 2 \expval{\cos^2\delta} +6 )$, and
$b = \tfrac {1}{10} ( 3 \expval{\cos^2\delta}-1 )$ where pointed braces refer to an ensemble value. From a single steady-state measurement, one cannot determine the extent to which nonzero $\expval{\cos^2\delta}$ arises by an intrinsic misalignment of the absorption and emission dipole moments or by rotational diffusion, however, time resolved measurements or steady-state measurements at various viscosities can be used to elucidate the relative proportions \cite{Lakowicz:2008aa}. Although we find $\delta$ to be more intuitive, $a$ and $b$ are more often cast in terms of the more abstract angle $\theta$, defined as the angle between the emission dipole and the azimuth of the excitation polarization \cite{Arteaga:2012ac,Lakowicz:2008aa}. In terms of this angle, $a = (1 + \expval{\cos^2\theta})/2$, and $b = (3 \expval{\cos^2\theta}-1)/4$. Another often used parameter is fluorescence anisotropy $r$, which is directly $r = 2b$ \cite{Arteaga:2012ac}, and can be obtained from the scattering matrix in \cref{eq:FIso} by
\begin{equation}\label{eq:FAfromM}
r = \frac{2(S_{22} - S_{21})}{3(S_{11}-S_{12})+S_{21}-S_{22}},
\end{equation}
which is independent of the scattering angle, and insensitive to uniform scaling of $\mathbf S$. In the trivial case of a medium that does not impart anisotropy on propagation at either the excitation or emission wavelengths, $\mathbf F\propto\mathbf S$, and thus \cref{eq:FAfromM} may be applied to the elements of $\mathbf F$ to obtain the fluorescence anisotropy, which is the only parameter for such a medium. 

Although it is possible for a medium that imparts general anisotropy on wave propagation to have an isotropic distribution of fluorophores, we will consider only the more simple case of an optically active medium (nonzero $A$, $C\!D$, and $C\!B$) at both the excitation and emission wavelengths. 
In this case, the Mueller matrix has the analytic form
\begin{multline}\label{eq:MOA}
\mathbf M(z) = \\
 e^{-A}\mqty[
  \cosh(C\!D) & 0 & 0 & \sinh (C\!D) \\
  0 & \cos (C\!B)  &  \sin (C\!B) & 0 \\
  0 & - \sin (C\!B) & \cos (C\!B) & 0 \\
  \sinh (C\!D) & 0 & 0 & \cosh(C\!D) ].
  \end{multline}
$\mathbf F$ is obtained by combining \cref{eq:MOA,eq:SzHomo,eq:diffMuellerJones,eq:FIso}. An analytical solution is readily obtained when $\phi=0$, and $C\!D_\mathrm{e}=0$ because the generators of $C\!B$ and $A$ commute with $\mathbf S$ in the forward scattering direction, and thus $\mathbf S$ can be moved to the left of the integral in \cref{eq:SzHomo}. Furthermore, because the generators of $A$, $C\!B$, and $C\!D$ commute, the law of exponentials can be used to solve the integral, giving
\begin{equation}\label{eq:FisoOA}
\mathbf F = \mathbf S\bigl(\exp(d\mathbf m_\mathrm{x}) - \exp(d\mathbf m_\mathrm{e})\bigr)(\mathbf m_\mathrm{x} - \mathbf m_\mathrm{e})^{-1}.
\end{equation}
Expanded equations are provided in \cref{app:2}. There has been much interest in probing $C\!D_\mathrm{x}$ by measuring the differential intensity of fluorescence upon excitation with left and right circularly polarized light, a quantity usually called fluorescence detected circular dichroism (FDCD). Circular dichroism spectropolarimeters actually measure the ratio $M_{14}/M_{11}$, which is approximately $C\!D$ when $C\!D$ is small. Likewise, FDCD polarimeters measure $F_{14}/F_{11}$, but this quantity is a more complicated function. However, in the limit that $A_\mathrm{e}\rightarrow0$, and $\exp(-A_\mathrm{x})\rightarrow0$ (that is, no absorption at the emission wavelength and total absorption at the excitation wavelength), then $F_{14}/F_{11}\rightarrow C\!D_\mathrm{x}/A_\mathrm{x} = g_\mathrm{abs}/2$ where $g_\mathrm{abs}$ is the so-called dissymmetry factor. This particular result is independent of the scattering angle and does not depend on fluorescence anisotropy.

\begin{figure}[tb]
\centering
\includegraphics[width=0.8\linewidth]{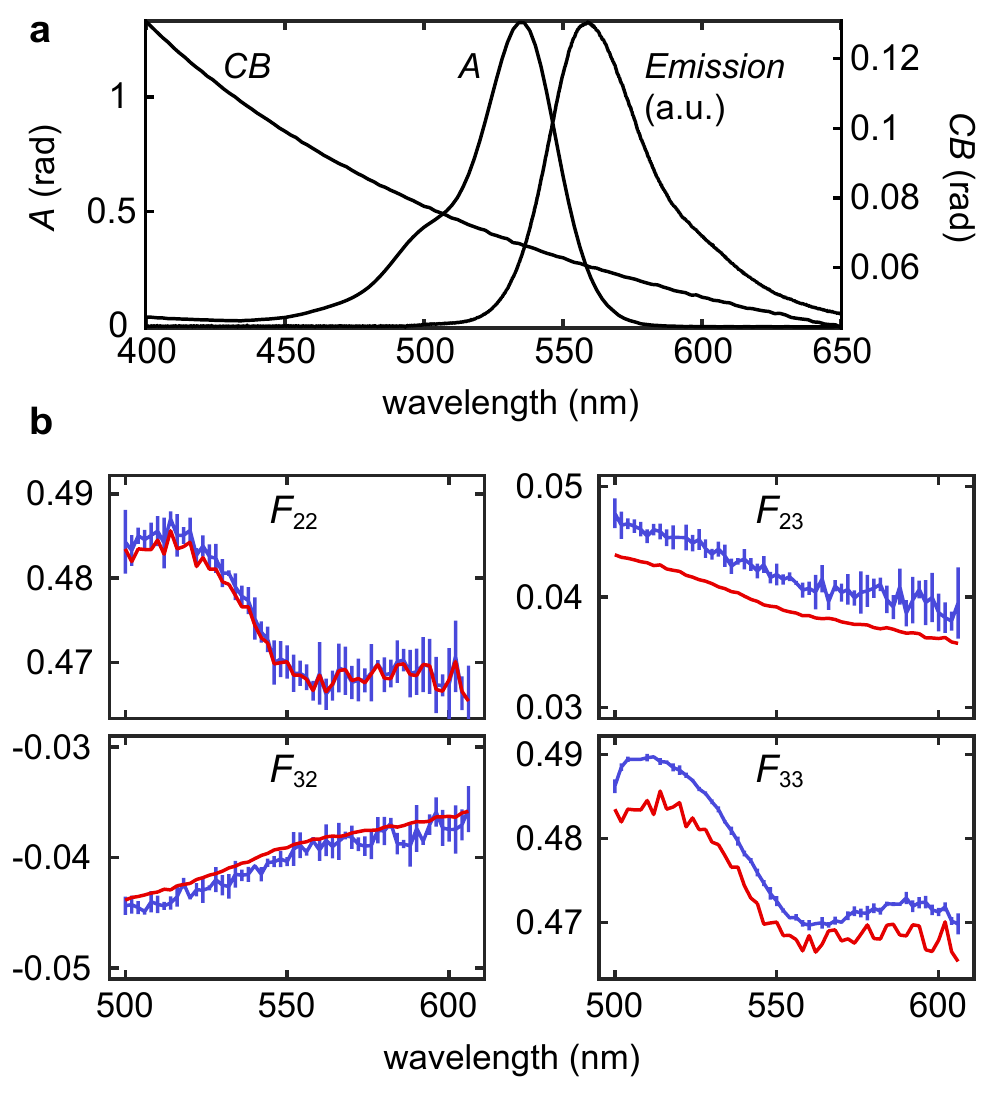}
\caption{Transmission and fluorescence polarization properties of $1\times10^{-4}$ M solution of rhodamine 6G in corn syrup. (a) Absorbance and $C\!B$ computed from the TMM. Data was not normalized so as to recover $A$. The emission spectrum is also given, in arbitrary units. (b) Red and blue spectra show, respectively, the measured and predicted FMM elements corresponding to the central block of the Mueller matrix; other elements are zero or one.}
\label{fig:OAsol}
\end{figure}

\begin{figure}[tb]
\centering
\includegraphics[width=\linewidth]{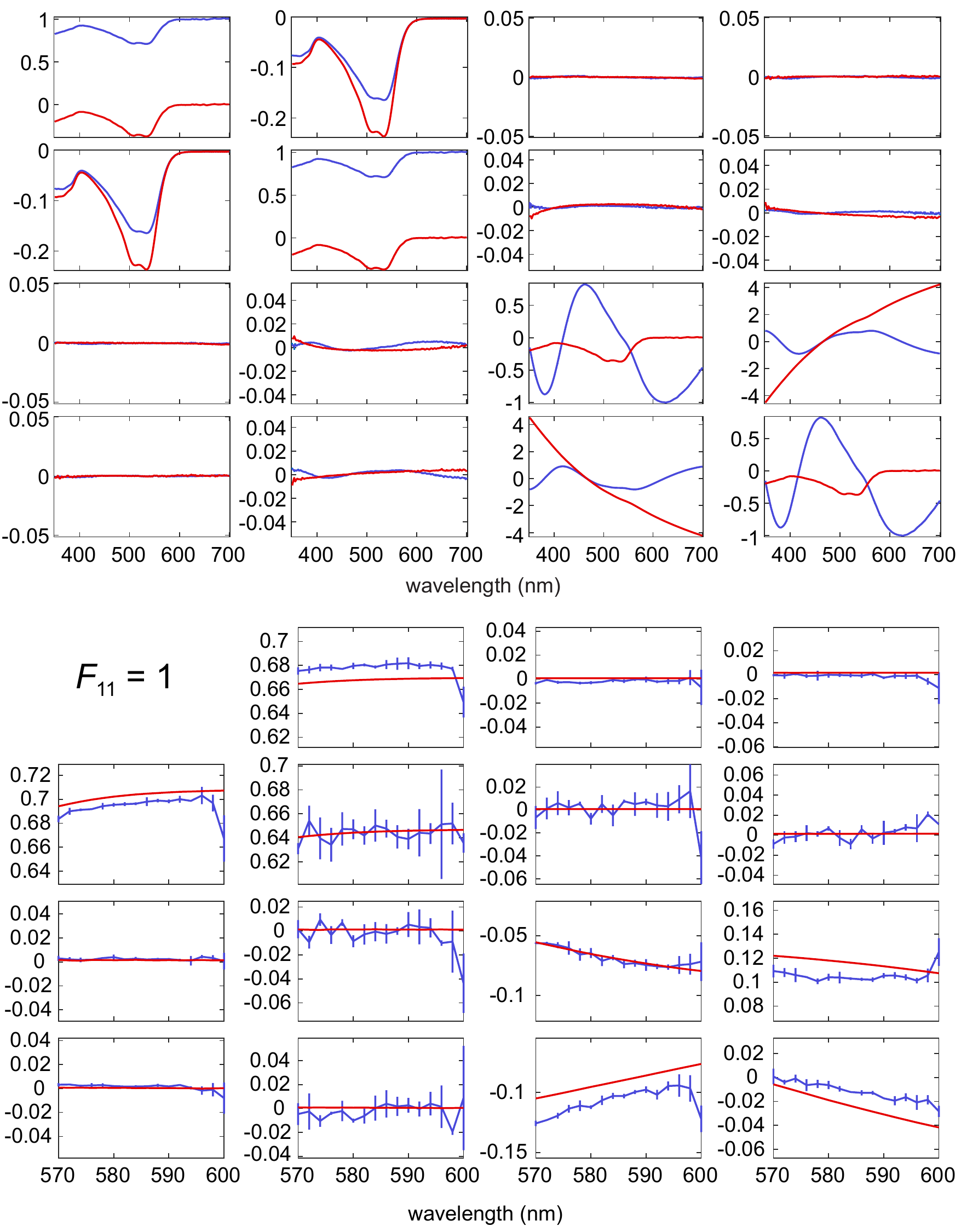}
\caption{(top) Blue and red spectra respectively show the measured TMM and the order-corrected differential TMM, $\mathbf m$.  (bottom) Blue and red spectra respectively show the measured and calculated fluorescence Mueller matrix, normalized by $F_{11}$. The calculation is based on transmission data and fitted values of the scattering matrix, $\mathbf S$.}
\label{fig:smn214237}
\end{figure}

\subsection{Dye in a viscous, optically active medium}

To test \cref{eq:FisoOA}, a $1\times10^{-4}$ M solution of rhodamine 6G in corn syrup (concentrated aqueous glucose) was prepared and placed in a 2 mm path length cuvette. The solution had small $C\!B$ at both the excitation and emission wavelengths but showed no $C\!D$ because rhodamine 6G, the sole absorbing molecule, is achiral. An unnormalized TMM measurement was used to recover $A$ and $C\!B$, as shown in \cref{fig:OAsol}a. The FMM emission spectrum was measured at $\phi=0$ using 441 nm excitation; the solution showed bright fluorescence despite being at the edge of the absorption band. Nonzero elements of the FMM, normalized by $F_{11}$, are shown in \cref{fig:OAsol}b along with predicted values that were calculated using equations in \cref{app:2} and the measured quantities in \cref{fig:OAsol}a. The calculation also requires knowing the fluorescence anisotropy. Because $F_{23}$ and $F_{32}$ are very small, we assumed that \cref{eq:FAfromM} could be used to a good approximation. The same line shape of $r$ was observed in non-optically active solutions of rhodamine 6G in glycerol, but with smaller values due to the lower viscosity of glycerol compared to corn syrup. Although propagation has a weak effect here, the $F_{23}$ and $F_{32}$ elements agree well with the expected values.

\subsection{Dye in an anisotropic polymer film}\label{sec:stretchedFilm}
This section demonstrates recovering $\mathbf S$ of a more general distribution of dye. Congo red (CR, 400 $\upmu$l of $1\times10^{-3}$ M) was added to polyvinyl alcohol (PVOH, 7.38 g in 80.6 g deionized H${_2}$O) and the mixture poured into a rectangular mold and left to dry in open air for several days. The resultant polymer was ca. 100 $\upmu$m thick after drying. The normalized TMM of the film was very close to the identity matrix, and the forward scattering FMM was ${\approx} \mathrm{diag}\mqty(1&0.5&0.5&0)$, which corresponds to a frozen isotropic distribution of dyes having collinear absorption and emission dipole moments. Unidirectional stretching of the film induced linear anisotropy, and induces a probability distribution that has a mean value along, and is symmetric about, the stretching direction. The stretching direction was oriented along the instrument $x$ axis by minimizing $L\!D'$ in transmission. By numerically integrating \cref{eq:Qformal} for such a distribution with its mean value along $x$, and evaluating \cref{eq:QtoS} for forward scattering, we determined that $\mathbf S$ in this case has the general form,
\begin{equation}\label{eq:SinStrechedFilm}
\mathbf S = \mqty[1&S_{12}&0&0\\S_{12}&S_{22}&0&0\\0&0&S_{33}&0\\0&0&0&0]
\end{equation}
where $\mathbf S$ has been normalized by $S_{11}$. This result agrees with others \cite{Tewarson:1967aa} as well as our intuition that $\mathbf S$ in this case should resemble the Mueller matrix of a partial polarizer with its transmission direction along $x$, except $S_{44}=0$ due to the incoherent nature of the scattering process. By this analogy, we can anticipate that all parameters in \cref{eq:SinStrechedFilm} will be positive numbers. 

\begin{figure}[tb!]
\centering
\includegraphics[width=\linewidth]{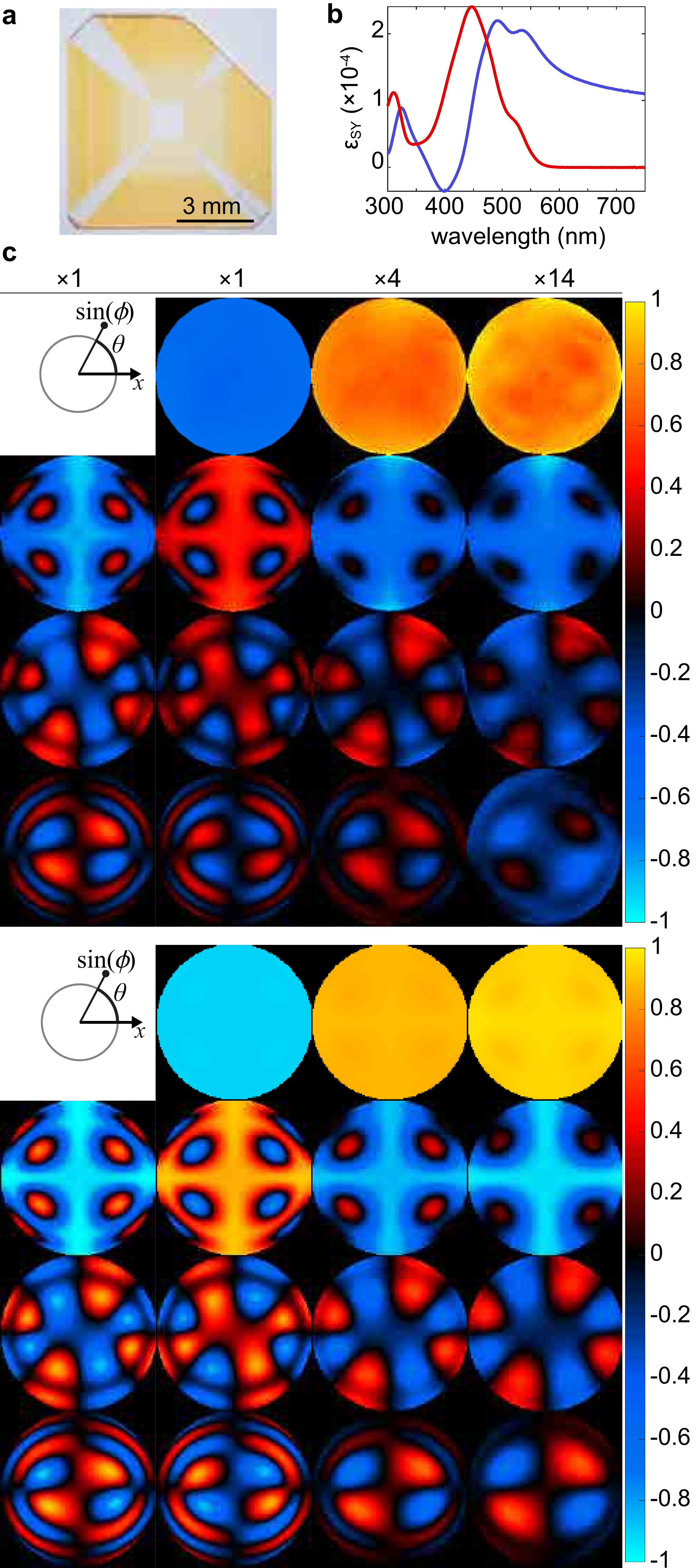}
\caption{(a) An (001) slice of a EDS/SY photographed in diffuse light. (b) Spectra of the (blue) real and (red) imaginary components of the dipolar term added to the electric permittivity tensor to account for SY in the crystal. (c) Comparison of the (top) measured and (bottom) calculated $\bm k$-space map of the FMM where multiplying factors for the columns of both maps are given at the top. The edge of the $\bm k$-space map corresponds to a scattering angle of 60 deg.}
\label{fig:syeds}
\end{figure}

Parameters $S_{12}$, $S_{22}$, and $S_{33}$ were obtained by fitting \cref{eq:SzHomo} to measured values of $\mathbf F$, and order-corrected values of $\mathbf m$ obtained from the measured TMM. The top panel of \cref{fig:smn214237} shows the TMM and $\mathbf m$ as a function of wavelength, from which $\mathbf m_\mathrm{x}$ and $\mathbf m_\mathrm{e}$ are obtained from their respective wavelengths. The blue lines on the bottom panel show the spectrum of $\mathbf F$, which was measured from 570 to 600 nm. Values were fitted using the Levenberg-Marquardt algorithm implemented in MATLAB. The fit was robust and insensitive to the initial values. Fitted values were $S_{12} = 0.727(1)$, $S_{22}=0.676(2)$, and $S_{33}=0.217(3)$, in which errors were estimated from the covariance matrix. \Cref{fig:smn214237} compares the measured FMM to one computed by \cref{eq:SzHomo} using the fitted parameters. The overall reduced $\chi^2$ figure of merit was 37, which is based on random errors obtained by repeated measurements. We attribute the large value to unaccounted for systematic errors, most likely originating from small birefringence in the focusing optics that were added to gather emission (see \cref{app:experimental}). Nonzero elements of $\mathbf S$ are approximately equal to the corresponding elements of the calculated FMM (red line in \cref{fig:smn214237}), excluding $S_{33}$, because elements $S_{12}$ and $S_{22}$ are more closely related to light emitted with states of linear polarization along the $\bm x$ and $\bm y$ axes, which are eigendirections for linear polarization. In contrast, $S_{33}$ is more related to light polarized along the $\pm 45^\circ$ axes, which are directions of maximum linear retardation. Thus, $S_{33}$ is more obscured by propagation effects than the other components of $\mathbf S$.

\subsection{Dye in a bianisotropic crystal}\label{sec:SYEDS}

\begin{figure}
\centering
\includegraphics[width=\linewidth]{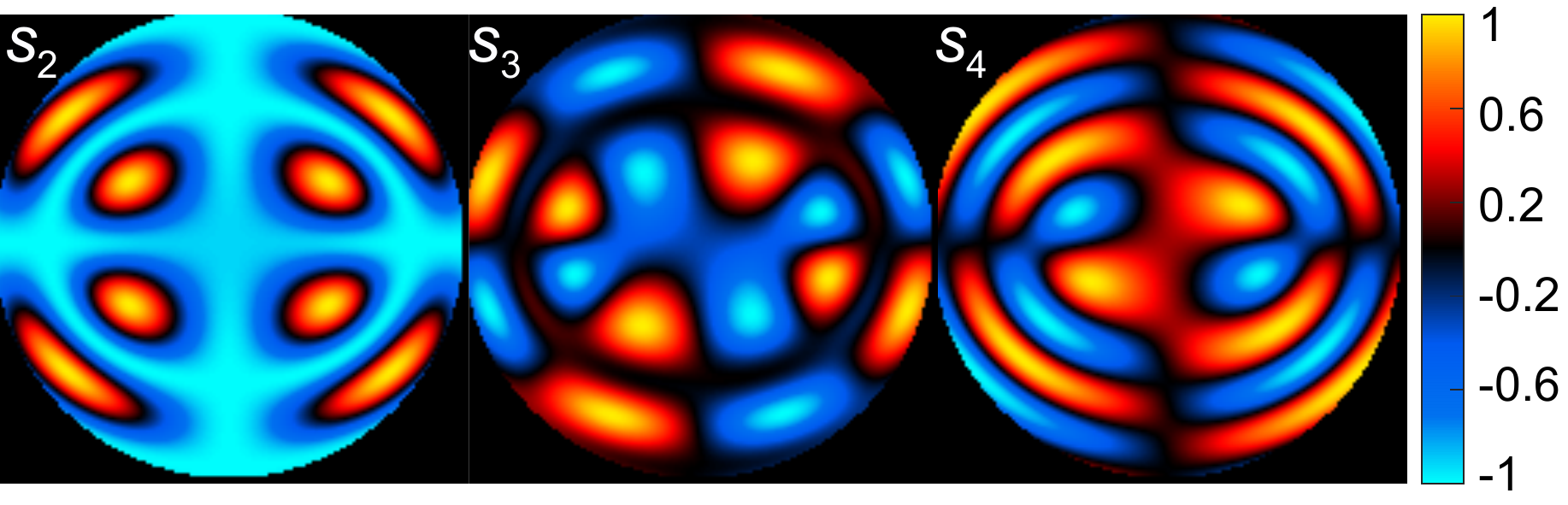}
\caption{Conoscopic maps of the last three elements of the normalized Stokes vector $\bm s$ corresponding to the matrix product $\bm s=\mathbf M_\mathrm{e}\bm s_0$ where $\mathbf M_\mathrm{e}$ is the TMM of EDS/SY computed with the partial wave method \cite{Nichols:2015aa}, and $\bm s_0$ is an incident Stokes vector that is linearly polarized with its azimuth of polarization aligned with the emission dipole moment of SY. All physical parameters for the crystal and the range of incident directions are identical to those used for the calculation in \cref{fig:syeds}c, in which the dye was aligned along $y$, thus $\bm s_0=\smqty[1& -1& 0& 0]^\mathrm{T}$ for all points in the conoscopic map. The elements $s_i$ are similar to the corresponding elements $F_i$ in the first column of the FMM shown at the bottom of \cref{fig:syeds}b.}
\label{fig:smn215908mtof}
\end{figure}

While molecules are only partially aligned within stretched films and membranes, growth impurities incorporated into the facets of crystals are often highly aligned \cite{Kahr:2001aa, Kahr:2016aa}. Impurities may incorporate homogeneously in space under suitable conditions, thereby providing an ideal medium in which to study the anisotropy of photo-physical properties \cite{Gurney:2000aa,Kaminsky:2003aa,Benedict:2006aa}. Such a medium is considered here. Crystals of ethylenediammonium sulfate (EDS) were grown from aqueous solution in the presence of sunset yellow FCF (SY), a disulfonated azo dye. EDS forms large, transparent, and optically clear crystals in the chiral, tetragonal point group 422 ($D_4$), and exhibits excellent (001) cleavage. We previously reported the optical functions of pure EDS \cite{Nichols:2016aa}. From polarized absorbance measurements of dyed crystals cut along different direction, we established that SY selectively incorporates into the \{102\} growth sectors and is oriented with its molecular plane perpendicular to the $a$ component of the growth direction. Dyes are highly aligned in that plane such that the dominant absorption band is polarized perpendicular to the four-fold ($c$) axis of EDS. Oblique incidence transmission and reflection Mueller matrix spectra could be accurately modeled by adding a single dipolar contribution to the materials electric permittivity tensor, which is plotted in \cref{fig:syeds}b. A fuller description of this material is given elsewhere \cite{Nichols:2018aa}.

We measured the FMM of this material using the $\bm k$-space imaging system described in \cref{app:experimental}. Collimated laser excitation at 450 nm was normally incident on a 683 $\upmu$m thick (001) section aligned such that the dye absorption direction was along the $y$ axis of the instrument. Forward scattered emission out to $\phi_\mathrm{e} = 60$ deg was gathered with a microscope objective, and the $\bm k$-space image focused onto a camera. The measured $\bm k$-space, i.e., conoscopic, Mueller matrix at $\lambda_\mathrm{e}=560$ nm is shown in the top panel of \cref{fig:syeds}c. Simulation of the measurement requires a model for $\mathbf S$. For totally aligned molecules, the integral in \cref{eq:Qformal} vanishes and thus the scattering can be represented by the Jones matrix
$\mathbf J_S = {\mqty[ \hat{\bm p}_\mathrm{e} &  \hat{\bm s}_\mathrm{e}]}^{\mathrm T}
		    \mathbf P
		    \mqty[ \hat{\bm p}_\mathrm{x} &  \hat{\bm s}_\mathrm{x}]$
from which it follows that $\mathbf S= \mathbf A(\mathbf J_S\otimes \mathbf J_S^*)\mathbf A^{-1}$. FMM measurements of SY in PVOH films revealed that the absorption and emission dipoles of SY are aligned, hence $\mathbf P = \hat{\bm u}_a\otimes\hat{\bm u}_\mathrm{x}$. With $\hat{\bm u}_\mathrm{x}$ oriented along $\bm y$,
\begin{equation}\label{eq:SforEDSSY}
\mathbf S \propto \mqty[1&-1&0&0 \\ -1&1&0&0 \\ 0&0&0&0 \\ 0&0&0&0],
\end{equation}
for any direction of $\bm k_\mathrm{e}$ in the global coordinate. Because we normalize the FMM by $F_{11}$, the proportionality in \cref{eq:SforEDSSY} may be changed to an equality without loss of generality. The FMM was simulated by combining this scattering matrix with \cref{eq:JwithBoundaries}, which describes propagation. The result is shown in the bottom panel of \cref{fig:syeds}c. Overall, the agreement with measurement is good, but we observed that the measured values are smaller, appearing  ``washed out'' compared to the simulation. This difference may arise from wavelength averaging as the emission bandpass filter had a window of 20 nm. Increasing the distribution of wavelengths in a measurement tends to dampen elements of the normalized Mueller matrix, which can be thought of as a loss of temporal coherence between propagating eigenwaves \cite{Nichols:2016ab}.

This data contains some clues as to how the TMM and FMM are related in the case of aligned dipoles. The rich patters in the elements of the FMM are more related to transmission properties at the emission wavelength. Indeed, the key features in the first column of the FMM can be captured by computing the conoscopic TMM at the emission wavelength \cite{Arteaga:2017ab, Nichols:2015aa} and multiplying the matrix by a linearly polarized Stokes vector with its azimuth of polarization aligned along the emission dipole moment of SY. The result of this calculation is shown in \cref{fig:smn215908mtof}, which closely resembles the first column of the FMM maps in \cref{fig:syeds}c. To a good approximation, the other columns of the FMM are related to the first column by overall scale factors that are given by the first row of the normalized TMM at the excitation wavelength. For this crystal, in which light was normally incident, the first row of $\mathbf M_\mathrm{x}$ is $\mqty[1 &   0.95 &  -0.22 &  -0.06]$, whereas the average values along the first row of the calculated FMM map in \cref{fig:syeds} are $\mqty[ 1 &   -0.91 &   0.22 &   0.06]$. We see that the scale factors for the second, third, and forth columns of the FMM are directly the negative of the corresponding elements in the first row of $\mathbf M_\mathrm{x}$. In this way, it is possible to approximate $\mathbf F$ without evaluating the integral in \cref{eq:SzHomo}.

\section{Conclusion}
To summarize, we have extended the differential Mueller matrix formalism by adding a source term that represents the Stokes vector of light generated within the medium by an incoherent scattering process. The general solution contains an intuitive integral term that propagates and sums newly generated light from all points along the propagation direction. This result was applied herein to fluorescence by connecting the source term to a scattering process driven by the Stokes vector at the excitation wavelength. An approximate approach to include boundaries was also given. Scattering and propagation were analyzed for several representative systems and the methods validated by measurement. While this work expressly considered fluorescence, all manner of luminescence or other incoherent scattering processes are encompassed within the general scope of this work. We expect these results to be illuminating to those researchers working in the newly emerging field of fluorescence polarimetry, as well as researchers involved in more classical fluorescence polarization analyses. With regards to possible future directions, this work provided analytical solutions to the integral equations for optically active isotropic media only, but it would be useful to have analytical solutions for general homogeneous media or other relevant media, such as continuously varying liquid crystalline phases, or elastically scattering media.

\appendix
\section{}\label{app:1}
Methods to obtain $\mathbf m$ are discussed here. The Jones formalism is used for convenience, in which a Mueller matrix given by $\mathbf M=\exp(\mathbf L)$, where $\mathbf m=\mathbf L/d$ with $\mathbf m$ having the form of \cref{eq:diffMuellerJones}, has the equivalent Jones matrix representation $\mathbf J = \exp(-i\mathbf N)$. The so-called differential Jones matrix $\mathbf N$ is \cite{Jones:1948aa}
\begin{equation}\label{eq:Nmatrix}
\mathbf N = \frac 12 \mqty[A+L & L'+iC \\ L'-iC & A-L],
\end{equation}
in which
$L = L\!B - iL\!D$, 
$L' = L\!B' - iL\!D$, 
$C = C\!B - iC\!D$. Equations will be given for the parameters of $\mathbf N$, from which $\mathbf m$ is obtained by $\mathbf m = \mathbf A(\mathbf N\oplus\mathbf N^*)\mathbf A^{-1}/d$, where $\oplus$ is the Kronecker sum and $\mathbf A$ is given in \cref{eq:Amatirx}. 

\subsection{Obtaining $\mathbf m$ from optical constants}\label{app:1b}
Homogeneous, non-magnetic, and Lorentz-reciprocal media can be described by an electric permittivity tensor $\bm\upepsilon$ and an optical activity tensor $\bm\upalpha$ that enter into the Tellegen constitutive relations \cite{Mackay:2010aa} $\bm D = \bm \upepsilon \bm E  -i\bm\upalpha \bm H$, and $\bm B = \bm H + i\bm\upalpha^{\textrm T} \bm E$, in Gaussian units. Light propagating in an unbounded medium in the direction of the unit vector $\hat{\bm k}$ accumulates
\begin{subequations}\label{XLfromU}\begin{gather}
A = -k_0d\Im\left(1/\sqrt{U_{11}}+1/\sqrt{U_{22}}\right) \\
L = k_0d\left(1/\sqrt{U_{11}}-1/\sqrt{U_{22}}\right)\\
\begin{split}
L' = k_0d\Bigl(1/\sqrt{(U_{11}+U_{22}+U_{21}+U_{12})/2}  \\ 
-1/\sqrt{(U_{11}+U_{22}-U_{21}-U_{12})/2}\Bigr)
\end{split} \\
C = k_0d(\hat{\bm k}^\mathrm{T}[\tr(\bm\upalpha)\mathbf I - \bm\upalpha]\hat{\bm k})
\end{gather}
over a distance $d$ where $k_0$ is the wavenumber, $\mathbf I$ is the identity matrix and,
$\mathbf U = {\mqty[ \hat{\bm p} &  \hat{\bm s}]}^{\mathrm T}
		    \mathbf \bm\upepsilon^{-1}
		    \mqty[ \hat{\bm p} &  \hat{\bm s}]$.
Unit vectors $\hat{\bm p}$ and $\bm{\hat s}$ are orthogonal directions perpendicular to $\bm k$ along which the electric field is projected. 
\end{subequations}


\begin{figure}
\centering
\includegraphics[width=1\linewidth]{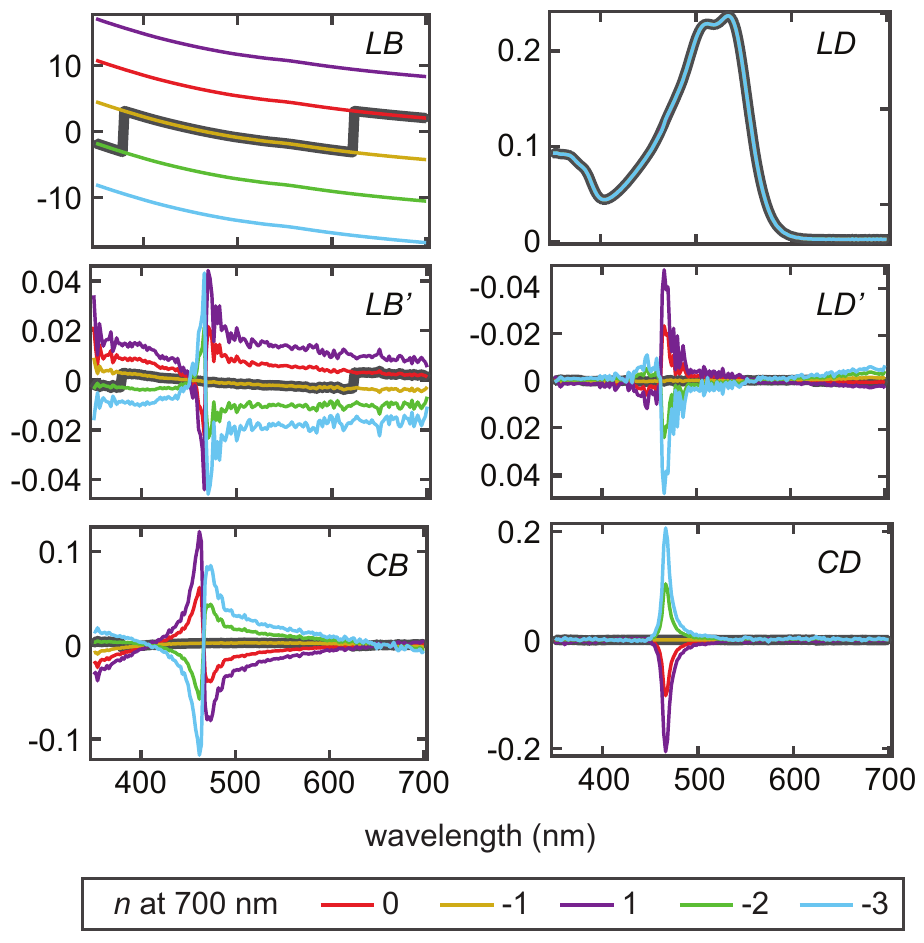}
\caption{Determination of the order of anisotropy for the TMM dataset used in \cref{sec:stretchedFilm}. The thick black line shows the parameters of $\mathbf m$ calculated from TMM data using \cref{JonesAnalyticInversion} with $n=0$ across the spectrum (no order correction). In each colored spectrum, $n$ is varied to maintain curve continuity, where the initial value of $n$ at 700 nm is given in the legend. All initial values of $n$ except $n=-1$ lead to artifacts that are most evident in $L\!D'$ and $C\!D$ spectra.}
\label{fig:orderChanges}
\end{figure}

\subsection{Obtaining $\mathbf m$ from a TMM measurement}
Assuming interfaces do not have a large effect, $\mathbf m$ can in principle be obtained directly from a TMM measurement by $\ln(\mathbf M)/d = \mathbf m$, but the physical solution must be selected from the generally infinite number of solutions to the matrix logarithm \cite{Devlaminck:2014aa}. By definition, $\mathbf m$ is unambiguously the matrix that satisfies $\mathrm{d}\mathbf M/\mathrm{d}z = \mathbf m\mathbf M$, but usually $\mathbf M$ cannot be measured as a function of the path length. Our approach is to compute various possible solutions, and use other available information to select the physical branch of the matrix logarithm.

Analytical solutions are straightforward if $\mathbf M$ can be represented by a Jones matrix. Because experimental data contains noise, $\mathbf J$ is usually estimated from $\mathbf M$ by matrix filtering \cite{Cloude:1990aa}, but formally we can write
\begin{subequations}\label{MuellerJones2Jones}
\begin{align}
	J_{11} &= \sqrt{(M_{11} + M_{12} + M_{21} + M_{22})/2} \\
	J_{12} &= \bigl(M_{13} + M_{23} - i(M_{14} + M_{24})\bigr)/2J_{11} \\
	J_{21} &= \bigl(M_{31} + M_{32} + i(M_{41} + M_{42})\bigr)/2J_{11} \\
	J_{22} &= \bigl(M_{33} + M_{44} + i(M_{43} - M_{34})\bigr)/2J_{11}.
\end{align}
\end{subequations}
From the elements of $\mathbf J$, the parameters of each possible solution to $\mathbf N$ are
\begin{subequations}\label{JonesAnalyticInversion}\begin{gather}
	A = -\Re(2\ln(1/K)) \\
	L = i\Omega(J_{11} - J_{22}) \\
	L' = i\Omega(J_{12} + J_{21}) \\
	C = \Omega(J_{12} - J_{21}) \\
	\intertext{where,}
	K = 1/\sqrt{\det(\mathbf J)} \\
	\label{JonesAnalyticInversion-arccos}
	T = 2\acos\bigl(K(J_{11} + J_{22})/2\bigr)  \\
	\label{JonesAnalyticInversion-XtraFactor}
	\Omega = K( T + 2\pi n)/(2\sin(T/2)),
\end{gather}\end{subequations}
which are equivalent to published relations \cite{Arteaga:2010aa} except that we have added the factor $2\pi n$ in \cref{JonesAnalyticInversion-XtraFactor} where $n$ is any integer. As anisotropy is uniformly increased from 0, the value of $n$ increments as $0,-1,1,-2,2\dotsc$. If $\mathbf M$ is measured spectroscopically with sufficiently high resolution, $n$ only needs to be known at one wavelength if $\mathbf m(\lambda)$ is assumed to be continuous. Our approach is to assign an initial value of $n$ at the longest measured wavelength and then use curve continuity to determine $n$ elsewhere. Comparing spectra for different initial values of $n$ can in some cases elucidate the physical value. Such a comparison is made in \cref{fig:orderChanges} using the TMM data in \cref{fig:smn214237}. All but the initial value of $n=-1$ lead to large values of $L\!D'$ and $C\!D$ around 466 nm that are unphysical. This approach should be used with caution, however. While it is robust for inverting simulated (i.e., perfect) Mueller matrices of absorbing media, even small amounts of added noise (0.1\%) can create instabilities in \cref{JonesAnalyticInversion} that generally grow with $n$. A numerical approach to order correction suggested by others is likewise very sensitive to measurement errors \cite{Devlaminck:2014aa}. Our polarimeter operates with systematic errors of $\approx 0.0005$ in transmission mode, and thus we were confident that our assignments of $\mathbf m$ were correct. Yet generally solving $\mathbf m=\ln(\mathbf M)$ for imperfect data with due regard for multivaluedness remains a challenge.

\section{}\label{app:2}
In a medium with nonzero $A_\mathrm{x}$, $A_\mathrm{e}$, $C\!B_\mathrm{x}$, $C\!B_\mathrm{e}$, and $C\!D_\mathrm{x}$, the FMM $\mathbf F$ in the forward scattering direction ($\phi=0$) has the nonvanishing elements,
\begin{widetext}
\begin{subequations}\begin{gather}
F_{11} = S_{11} \frac{ (A_\mathrm{x} - A_\mathrm{e})(e^{-A_\mathrm{e}}-e^{-A_\mathrm{x}}\cosh C\!D_\mathrm{x})- C\!D_x e^{-A_\mathrm{x}}\sinh C\!D_x}
	{(A_\mathrm{x} - A_\mathrm{e})^2 - C\!D_\mathrm{x}^2}\\
F_{14} = S_{11} \frac{ (A_\mathrm{e} - A_\mathrm{x})e^{-A_\mathrm{x}}\sinh C\!D_x+ C\!D_x (e^{-A_\mathrm{e}}-e^{-A_\mathrm{x}}\cosh C\!D_\mathrm{x})}
	{(A_\mathrm{x} - A_\mathrm{e})^2 - C\!D_\mathrm{x}^2}\\
F_{22} = F_{33} = S_{22} \frac { (A_\mathrm{x} - A_\mathrm{e})(e^{-A_\mathrm{e}}\cos C\!B_\mathrm{e}  - e^{-A_\mathrm{x}}\cos C\!B_\mathrm{x})  +
                     (C\!B_\mathrm{x} - C\!B_\mathrm{e})(e^{-A_\mathrm{x}}\sin C\!B_\mathrm{x}  - e^{-A_\mathrm{e}}\sin C\!B_\mathrm{e}) }
                     {(A_\mathrm{x} - A_\mathrm{e})^2 + (C\!B_\mathrm{x} - C\!B_\mathrm{e})^2 } \\
F_{23} = -F_{23} = S_{22} \frac {  (C\!B_\mathrm{x} - C\!B_\mathrm{e})(e^{-A_\mathrm{e}}\cos C\!B_\mathrm{e}  - e^{-A_\mathrm{x}}\cos C\!B_\mathrm{x}) +
                      (A_\mathrm{x} - A_\mathrm{e})(e^{-A_\mathrm{e}}\sin C\!B_\mathrm{e}  - e^{-A_\mathrm{x}}\sin C\!B_\mathrm{x})}
                     {(A_\mathrm{x} - A_\mathrm{e})^2 + (C\!B_\mathrm{x} - C\!B_\mathrm{e})^2 }
\end{gather}\end{subequations}
\end{widetext}
in which the scattering matrix elements $\mathbf S_{ii}$ are given by \cref{eq:FIso}. 

\begin{acknowledgments}
This work was supported by the National Science Foundation (DMR-1105000), the National Institutes of Health (5R21GM107774-02), a Margaret Strauss Kramer Fellowship and a Margaret and Herman Sokol Fellowship from the NYU Department of Chemistry to ATM.
\end{acknowledgments}

\bibliography{Fluorescence2017_2}

\end{document}